\documentstyle[aps,prb,twocolumn,graphics]{revtex}

\begin{document}

\title{The probability distribution of the conductance in anisotropic systems}
\author{Marc R{\"{u}}hl{\"{a}}nder$^1$, Peter Marko{\v{s}}$^{1,2}$
and C. M. Soukoulis$^1$ \\
$^1$Ames Laboratory and Department of Physics and Astronomy, 
Iowa State University, Ames, Iowa 50012\\
$^2$Institute of Physics, Slovak Academy of Sciences,
D{\'{u}}bravsk{\'{a}} cesta 9, 842 28 Bratislava, Slovakia}
\maketitle

\begin{abstract}
We investigate the probability distribution $p(g)$ of the conductance $g$
in anisotropic two--dimensional systems. The scaling procedure applicable
to mapping the conductance distributions of localized anisotropic systems
to the corresponding isotropic one can be extended to systems at the
critical point of the metal--to--insulator transition.
Instead of the squares used for isotropic systems, one should use rectangles
for the anisotropic ones. At the critical point, the ratio of the side lengths 
must be equal to the squre root of the ratio of the critical values of the 
quasi--one--dimensional scaling functions. For localized systems,
the ratio of the side lengths must be equal to the ratio of the localization
lengths.
\end{abstract}

The presence of disorder\cite{Sou99b} may allow a system to make a
transition from metallic to insulating behaviour by varying the
Fermi energy in an energy range where both extended and localized
states are found, separated by a mobility edge. Characterizing this
transition, one can employ transport properties, such as the conductance,
or properties of the system's eigenstates, such as the correlation
length for extended, metallic states or the localization
length $\xi$ for insulating states. At the mobility edge, a determination
of the complete probability distribution $p(g)$ of the conductance $g$
(in units of $e^2/h$) is needed. The critical point of the transition from 
metallic states to Anderson localized ones\cite{Krm93}
is of particular interest. The distributions are well--known to be
normal and log--normal off the mobility edge towards the extended and 
the localized regime respectively, whereas the exact form of the
critical distribution is still under 
investigation.\cite{Sle97,Sou99a,Bra01,Sha90,Mks99,Mtt99,Wan98b,Ple98}
For example, contrary to expectations, the critical distribution seems
to vary even within the same universality class, depending on the
boundary conditions perpendicular to the direction in which transport
occurs.\cite{Sle97,Sou99a,Bra01} Also, questions about the exact form 
of the large--$g$ tail ($g > 1$) remain unanswered. Where calculations in 
$2 + \varepsilon$ dimensions\cite{Sha90} indicate higher cumulants
to diverge with system size, leading to a power law tail, numerical
calculations\cite{Mks99} in three dimensions and analytical results
for quasi--one--dimensional wires\cite{Bra01} show an exponential
decay. 

Anisotropic systems have recently been the focus of particular
attention.\cite{LiQ97,Mil97,Zam96,Dup97,Ruh01,Evg01} It is generally
accepted that anisotropy does not change the universality class and
that isotropic results can be recovered by performing a proper
scaling of the anisotropic results. For anisotropic systems in a
localized state, it is reasonable to assume that scaling the dimensions
of the system by the corresponding localization lengths will make
the system effectively isotropic. This procedure has been applied
successfully\cite{LiQ97} to the scaling function $\Lambda = \lambda_M/M$,
which is a function of $\xi/M$, where $\lambda_M$ denotes the finite
size localization length of a quasi--one--dimensional strip of finite
width $M$ and $\lambda_M \to \xi$ as $M \to \infty$. It was also
shown\cite{Wan97} that the same scaling procedure works for the probability
distribution in such a system.

In order to test the approach for critical states one must
either face the numerical challenge of large three--dimensional systems
or take into account additional interactions (beyond the disorder
potential) such as spin--orbit coupling.\cite{Hik80} Another possibility 
is the introduction of external magnetic fields as e.g.\ in integer quantum
Hall systems\cite{Huc95} or tight binding models with random magnetic
flux.\cite{Fur99} However, as a result of Anderson localization,
extended states do not exist in two--dimensional systems of
non--interacting electrons in a magnetic field, except at a singular
energy near the center of each of the Landau subbands. At these
critical energies $E_c$ the localization length $\xi$ diverges
with a critical exponent $\nu$: $\xi \propto |E - E_c|^{-\nu}$.

Because significant finite size effects have to be expected,
we decided to concentrate our research on two--dimensional systems
although the exact form of the critical distribution of the
conductance depends on the dimensionality of the system.\cite{Mks99}
The investigation of the self--averaging quantity $\xi$ in integer
quantum Hall systems yielded very encouraging results,\cite{Ruh01}
supporting the expectation that quantities of anisotropic systems
can indeed be mapped to isotropic values by a simple rescaling scheme.

In this paper, we show a method of mapping the probability distributions
of the conductance of anisotropic two--dimensional systems with a
magnetic field perpendicular to the plane or with spin--orbit
coupling to the probability distribution of the conductance for the
corresponding isotropic system at the critical point, using a tight
binding model. It turns out that the ratio of the squares of the
side lengths $L_x$, $L_y$ of the anisotropic system should be chosen
equal to the ratio of the critical values $\Lambda^c_x$, $\Lambda^c_y$ 
of the quasi--one--dimensional scaling functions:

\begin{equation}
\frac{L_x^2}{L_y^2} = \frac{\Lambda^c_x}{\Lambda^c_y}
\end{equation}

\noindent
In the following, we first describe the models and the numerical method we
employed. Then we present and discuss our numerical results
and finally summarize the conclusions of this work.

The tight binding model uses the Hamiltonian

\begin{equation}
{\mathcal H} = \sum_{n,\tau} \left| n\tau \right\rangle \varepsilon_n
\left\langle n\tau \right| + \sum _{n,\tau,n',\tau '} \left| n\tau 
\right\rangle V_{n,n'} \left\langle n'\tau ' \right|
\end{equation}

\noindent
where $n,n'$ denotes the lattice site. Without spin--orbit interaction
the ``variables'' $\tau,\tau '$ take on only one value and the hopping
integrals $V_{n,n'}$ are scalar, otherwise they are $2 \times 2$ matrices
and the spin variables take on the values $1$ or $-1$. In either case
the site energies $\varepsilon_n$ are independent of $\tau$ and we take
into account interactions only between neighbouring lattice sites.

An external magnetic field enters the Hamiltonian via its vector
potential $\mathbf A$ (${\mathbf \nabla} \times {\mathbf A} = {\mathbf B}$),
which appears in the phases of the hopping integrals:

\begin{equation}
V_{n,n'} = t_{n,n'}^0 e^{-2\pi i(e/h)
\int_{{\mathbf r}_n}^{{\mathbf r}_{n'}} {\mathbf A}({\mathbf r})
{\mathrm d}{\mathbf r}}
\end{equation}

\noindent
The integral connects the lattice sites $n$ (at ${\mathbf r}_n$) and
$n'$ (at ${\mathbf r}_{n'}$) in a straight line. For the sytems under
consideration, where the magnetic induction $\mathbf B$ is perpendicular
to the plane of the two--dimensional lattice, the gauge for the vector
potential can be chosen such that the phases vanish in the direction
perpendicular to $\mathbf A$ and are integer multiples of some number
$2\pi\alpha$ in the direction parallel $\mathbf A$. The value of the
parameter $\alpha$ then completely characterizes the influences of
the magnetic field on the system. For rational $\alpha$, the denominator
determines the number of bands in the density of states of the system
without disorder.

The Evangelou--Ziman model\cite{Eva87} incorporates spin--orbit coupling
by using the following hopping integrals:

\begin{equation}
V_{n,n'}^{\tau ,\tau '} = t_{n,n'}^0 \left[ \delta_{\tau,\tau '}
+ \mu i \sum_\nu \sigma_{\tau, \tau '}^\nu t_{n,n'}^\nu \right]
\end{equation}

\noindent
where $\nu = x,y,z$ and $\sigma^\nu$ are the Pauli matrices. The parameter
$\mu$ characterizes the strength of the spin--orbit interaction.

Both systems may be made anisotropic by chosing the value of $t^0$
to be different in the two directions within the plane. Otherwise
this parameter is a constant, independent of lattice site $n$.
We bring disorder to the system by chosing all the site energies
independently from a rectangular distribution of width $W$ centered at $0$,
so that $W$ is a measure of the strength of the disorder. The parameters
$t^\nu$ are also randomly selected from a uniform distribution on
$[-1/2,1/2]$. The energy scale is set by the larger of the two values
for $t^0$, which is taken to be unity.

We calculate the conductance from the Landauer formula\cite{Ldr57}

\begin{equation}
g = \mbox{Tr} \left( {\mathrm t}^\dagger {\mathrm t} \right)
\end{equation}

\noindent
where $\mathrm t$ is the transmission matrix. We suppose two semi--infinte
leads are attached to opposite sides of the sample. Then $\mathrm t$ determines
the transmission of an electron through the sample.  The numerical
procedure is based on the algorithms published by Ando and Pendry.\cite{And91}

The critical conductance distributions we calculated for isotropic
quantum Hall systems at different disorder strengths show that finite
size effects become stronger the weaker the disorder. Where systems
with $W = 4.0$ and $W = 2.0$ show a basically size--independent critical
distribution of the conductance for squares of $64 \times 64$ lattice sites,
at $W = 0.5$ finite size effects are still somewhat noticeable up to
systems with $192 \times 192$ lattice sites.
The anisotropic quantum Hall systems we investigated are
characterized by $\alpha = \frac{1}{8}$.
The anisotropies we chose were $t^0_x/t^0_y = 0.5$ at $W = 0.5$ and
$t^0_x/t^0_y = 0.8$ at $W = 0.1$. The latter was chosen mainly because
we already had the data for the quasi--one--dimensional scaling function.
As the disorder is even weaker than in the first case, finite size effects
are even stronger, and even at $240 \times 240$ lattice sites the
conductance distribution is far from the one we expect from our calculations
of isotropic systems.
Therefore, we will not be able to show that our procedure maps the
two anisotropic conductance distributions to the critical distribution
of isotropic systems for this extreme case. We will however be able to
prove the somewhat weaker claim that our method transforms the two anisotropic
distributions so that both have the same shape. In a square system, one
expects that the distribution in the difficult hopping direction
shows a more localized character than the one in the easy hopping direction.
In an isotropic system, the distribution obviously cannot depend on the 
direction of transport. By making the system rectangular rather than
square, i.e.\ shorter in the difficult hopping direction, it should be
possible to obtain distributions in the two directions that are the
same, thus making the anisotropic system effectively behave isotropically.

The task now is ``How do we choose the correct ratio of side lengths
of the rectangle?'' From the research on localized systems\cite{LiQ97,Wan97}
we know that in those cases, the ratio should be equal to the ratio of the
localization lengths:

\begin{equation}
\label{exact}
\left( \frac{L_x}{\xi_x} = \frac{L_y}{\xi_y} \qquad \Longrightarrow \right)
\qquad \frac{L_x}{L_y} = \frac{\xi_x}{\xi_y} 
\end{equation}

\noindent
for localized systems,
as these are obviously the appropriate length scales in their
respective directions.
This is of no use for critical systems as both localization lengths diverge
at the transition. A closely corresponding {\it non}--diverging quantity
is, however, available in the scaling function $\Lambda_M = \lambda_M/M$,
which has a finite critical value, independent of the system width
$M$. The finite size localization lengths $\lambda_{M,x}$ and 
$\lambda_{M,y}$ have $\xi_x$ and $\xi_y$ respectively as their
large--$M$ limits, and for big enough systems we can approximate
Eq.\ \ref{exact} by

\begin{equation}
\label{almost}
\frac{L_x}{L_y} = \frac{\lambda_{M,x}}{\lambda_{M,y}} \qquad 
\end{equation}

\noindent
for localized systems at ``large enough'' $M$.
The meaning of $M$ in this context would be that of the system width
perpendicular to the direction in which the localization length is
measured, i.e.\ $M = L_y$ in $\lambda_{M,x}$ and $M = L_x$ in
$\lambda_{M,y}$. Now, by multiplying both sides of Eq.\ \ref{almost}
by $L_x/L_y$ we have

\begin{equation}
\frac{L_x^2}{L_y^2} = \frac{L_x}{\lambda_{M,y}} \cdot 
\frac{\lambda_{M,x}}{L_y} = \frac{\Lambda_{M,x}}{\Lambda_{M,y}}
\end{equation}

\noindent
for ``large'' localized systems.
Now $\Lambda_M$ is a continuous function of $E$ and for ``large
enough'' systems at $E_c$ should have reached its critical value.
Therefore, we arrive at the conclusion that

\begin{equation}
\label{root}
\frac{L_x}{L_y} = \sqrt{\frac{\Lambda_x^c}{\Lambda_y^c}}
\end{equation}

\noindent
should be the correct ratio for critical systems in order to make
them behave like isotropic ones. Noting the 
relationship\cite{Wan98b,Zam96,Wan97}

\begin{equation}
\sqrt{\Lambda_x^c \cdot \Lambda_y^c} = \Lambda_{\mbox{iso}}^c
\end{equation}

\noindent
we can write Eq.\ \ref{root} in the alternate form

\begin{equation}
\frac{L_x}{L_y} = \frac{\Lambda_x^c}{\Lambda_{\mbox{iso}}^c}
= \frac{\Lambda_{\mbox{iso}}^c}{\Lambda_y^c}.
\end{equation}

\begin{figure}[h]
\resizebox{3.0in}{3.0in}{ \includegraphics{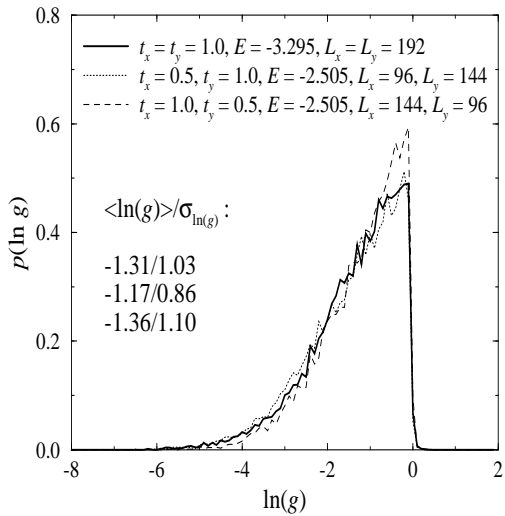} }
\caption{\label{cond05} The conductance distributions of an anisotropic
rectangular system with a ratio of side lengths chosen according to
Eq.\ \ref{root}. For comparison the corresponding distribution of an
isotropic system is shown as well.}
\end{figure}

\begin{figure}[h]
\resizebox{3.0in}{3.5in}{ \includegraphics{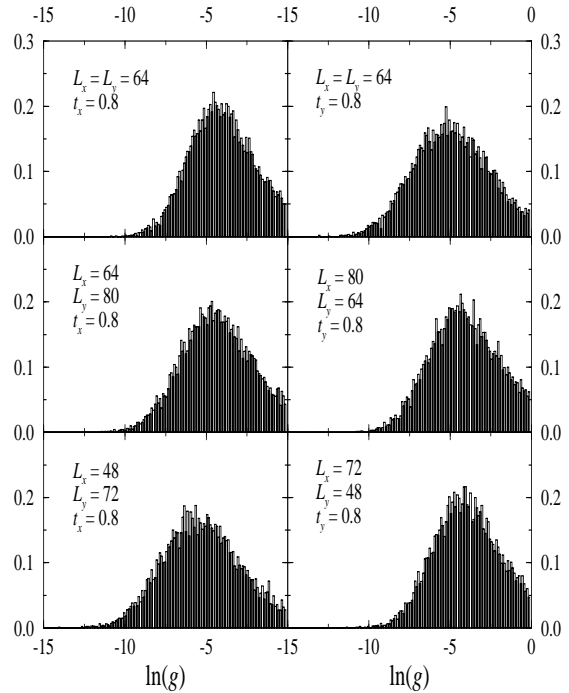} }
\caption{\label{cond08} Conductance distributions of an anisotropic system
for varying ratios of the side lengths. The left panels refer to transport
in the easy hopping direction, the right panels to transport in the
difficult hopping direction. The ratio of side lenghts is $1.0$ for the
top row, $1.25$ for the middle row and $1.5$ for the bottom row.}
\end{figure}

We tested this prediction on the system with $t^0_x/t^0_y = 0.5$
and $W = 0.5$, where the ratio $\sqrt{\Lambda_x^c/\Lambda_y^c}$
is roughly $1.5$. The result is shown in Fig.\ \ref{cond05}
together with the critical distribution for the isotropic system.
The agreement is very good. For the other system with $t^0_x/t^0_y = 0.8$
and $W = 0.8$, we have to deal with stronger finite size effects
and cannot expect to approach the form of the critical distribution
we see in Figs.\ \ref{cond05} for reasonable system
sizes. Instead we merely show in Fig.\ \ref{cond08} how the critical
distributions change with the ratio of side lengths. The best value
for the ratio according to Eq.\ \ref{root} would be roughly $1.23$.
Fig.\ \ref{cond08} shows results for ratios of $1.0$, $1.25$ and
$1.5$. The averages of $\ln(g)$ for the easy hopping direction
decrease with increasing ratio from $-4.09$ for the square to
$-4.40$ and $-5.30$, while the averages for the difficult hopping
direction increase from $-4.94$ for the square to $-4.18$ and $-4.03$.
Similarly, the standard deviations increase for the easy hopping direction
from $1.90$ for the square to $2.07$ and $2.30$, while they decrease
for the difficult hopping direction from $2.24$ for the square to $2.01$ 
and $1.91$. The values for a ratio of $1.25$ are not equal but
reasonably close, so that for larger systems, where a ratio of $1.23$
might be practicable, we expect a better agreement of the two
probability distributions.

Taking the best--ratio rectangle as the ``undeformed'' base, we
can also see from Fig.\ \ref{cond08} that similar ``deformations''
have similar effects in the two directions, that is, reducing the
ratio by a factor $\gamma$ in one direction will cause the ensemble
average $<\ln(g)>$ in that direction to increase, and the standard
deviation to decrease, while the trend is opposite in the perpendicular
direction. However, by reducing the ratio by the same factor $\gamma$
in the other direction, will result roughly in the same distributions
as before, but with the one associated with the easy direction before
now assigned to the difficult direction and {\it vice versa}.

\begin{figure}[h]
\resizebox{3.0in}{3.0in}{ \includegraphics{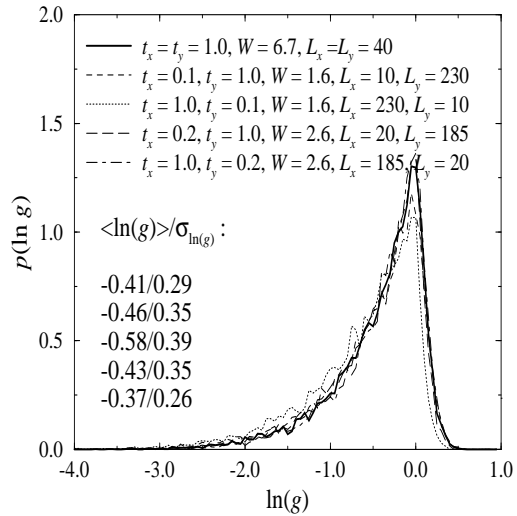} }
\caption{\label{spin} The conductance distributions for an isotropic
and two anisotropic systems with spin--orbit coupling.}
\end{figure}

That the same procedure also works for sytems with spin--orbit
coupling is shown in Fig.\ \ref{spin}, where we plot the conductance
distribution for an isotropic system with $\mu = 1.0$ at $E = 0.1$
and $W_c = 6.7$ together with that of two anisotropic systems, one with
$\mu = 1.0$ and $t^0_x/t^0_y = 0.1$ at $E = 0.1$ and $W_c = 1.6$, the other
with $\mu = 1.0$ and $t^0_x/t^0_y = 0.2$ at $E = 0.1$ and $W_c = 2.6$.
The ratio of sidelengths, according to Eq.\ \ref{root}, should be
$23.0$ for the latter. We chose $40 \times 40$ lattice sites for
the isotropic system, $10 \times 230$ lattice sites for the
strongly anisotropic one and $20 \times 185$ lattice sites for the
anisotropic system with the weaker anisotropy. 
Again the agreement is very good.

We have shown that the scaling procedure applicable to mapping
the conductance distributions of localized anisotropic systems
to the corresponding isotropic one can be extended in a straightforward
manner to systems at the critical point of the Anderson 
localization--delocalization transition in both unitary and
symplectic two--dimensional systems. Instead of the squares used
for isotropic systems, one should use rectangles for the anisotropic
ones, with a ratio of side lengths equal to the square root of
the ratio of the critical values of the quasi--one--dimensional scaling
function.

Ames Laboratory is operated for the U.S.\ Department of Energy by
Iowa State University under Contract No.\ W--7405--Eng--82.
This work was supported by the Director for Energy Research,
Office of Basic Science. P.M.\ would like to thank IITAP and Ames Laboratory at
Iowa State University for their hospitality and support and the Slovak
Grant Agency for financial support.

\end{document}